\documentclass[pra,reprint,longbibliography,superscriptaddress,aps%,groupedaddress]]
]{revtex4-1}

\usepackage{graphicx}
\usepackage{amsmath}
\usepackage{amssymb}
\usepackage{amsfonts}
\usepackage{epsfig}
\usepackage{dcolumn}
\usepackage{mathtools}
\begin{document}

\title{Cross-Phase Modulation Enhancement Via a Resonating Cavity: Semiclassical Description}

\author{Juli\'{a}n Mart\'{i}nez-Rinc\'{o}n}
\email{jmarti41@ur.rochester.edu}
\affiliation{Department of Physics and Astronomy, University of Rochester, Rochester, New York 14627, USA}
\affiliation{Center for Coherence and Quantum Optics, University of Rochester, Rochester, New York 14627, USA}
\author{John C. Howell}
\affiliation{Department of Physics and Astronomy, University of Rochester, Rochester, New York 14627, USA}
\affiliation{Center for Coherence and Quantum Optics, University of Rochester, Rochester, New York 14627, USA}
\affiliation{Institute of Optics, University of Rochester, Rochester, New York 14627, USA}
\affiliation{Institute for Quantum Studies, Chapman University, Orange, California 92866, USA}

%\author[1,2,*]{Juli\'{a}n Mart\'{i}nez-Rinc\'{o}n}
%\author[1,2,3,4]{John C. Howell}
%
%\affil[1]{Department of Physics and Astronomy, University of Rochester, Rochester, New York 14627}
%\affil[2]{Center for Coherence and Quantum Optics, University of Rochester, Rochester, New York 14627}
%\affil[3]{Institute of Optics, University of Rochester, Rochester, New York 14627}
%\affil[4]{Institute for Quantum Studies, Chapman University, 1 University Drive, Orange, California 92866}
%\affil[*]{Corresponding author: jmarti41@ur.rochester.edu}

\date{\today}

%\ociscodes{(190.0190) Nonlinear optics; (190.3270) Kerr effect; (140.4780) Optical Resonators; (000.1600) Classical and quantum physics; (020.1335) Atom optics.}

%\doi{\url{http://dx.doi.org/10.1364/ao.XX.XXXXXX}}

\begin{abstract}
We evaluate the advantages of performing cross-phase modulation (XPM) on a very-far-off-resonance atomic system. We consider a ladder system with a weak (few-photon level) control coherent field imparting a conditional nonlinear phase shift on a probe beam. We find that by coupling to an optical resonator the optimal XPM is enhanced proportional to the finesse of the resonator by a factor of $F/4\pi$. We present a semi-classical description of the system and show that the phenomenon is optimal in the self-defined condition of off-resonance-effective-cooperativity equal to one.   
\end{abstract}

%\setboolean{displaycopyright}{true}

%\begin{document}

\maketitle
%\thispagestyle{fancy}

%\ifthenelse{\boolean{shortarticle}}{\ifthenelse{\boolean{singlecolumn}}{\abscontentformatted}{\abscontent}}{}

\section{Introduction}

The possibility of affecting the phase of laser light with another one of different wavelength, or Cross-Phase Modulation (XPM), has been an engaging approach towards technological implementations due to its non-linear response. Coherent and strong light-light interaction is an ongoing fundamental goal for quantum processing of information \cite{ThesisFeizpour}.

%Cavity Quantum Electrodynamics (cavity QED), which describes a system composed of one two-level atom coupled to a single mode of an optical resonator, has been the building block for a vast variety of approaches in the ongoing goal of attaining full quantum control at the single atom and photon level. The primary requirement of such approaches is to satisfy the atom-photon strong-coupling regime at the single-photon level \cite{Kimble1998,Kimble2005}. This regime is defined by a large on-resonance cooperativity value, namely $\eta>1$. A cooperativity exceeding unity is normally understood to be when quantum phenomena play an important role in the system's dynamics \cite{RabiSplitting}. Strong photon-photon nonlinearities can be obtained because one photon is able to saturate the atomic response, and coherent control overcomes photon leaking out of the resonator and spontaneous emission \cite{PhotonBlockade}. Nevertheless, it was recently shown that many effects in multi-atoms cavity QED can be understood from a fully classical description, even within the strong-coupling regime \cite{ClassicalCavity}.

The first demonstration using cavity QED (Cavity Quantum Electrodynamics) to perform a controlled-phase shift was done twenty years ago \cite{Turchette}. A weak control beam induced a Kerr-type nonlinear phase shift in a probe beam using the birefringence of a single atom coupled to a high-finesse optical resonator. The experiment was performed in the \textit{bad cavity regime} and high absorption of the control-beam made it non-deterministic. The reported result of 0.28 rad conditioned phase shift per average intracavity control photon remained as the record for cavity-type XPM for many years~\cite{KristiXPM}. 

An alternative approach to XPM, based on EIT, uses an additional strong pump beam coupled to the separate atomic transitions of the control and probe beams \cite{Schmidt1996}. The advantage of this approach is a larger nonlinear atomic response with low absorption of the control beam. Much progress has been done following variants of this path \cite{LukinVelocities,Lukin2001,Zeeman,polarization,Zhu2003,Harris2003,Tripod,polarization2,slowpulses,slowpulses2,interference,slowpulses3}, nevertheless the largest phase shift to date, known to the authors of this work, on a probe pulse modulated by a control pulse of about $\sim$400 photons is of only 5 mrad ($\sim$12.5 $\mu$rad/photon) \cite{Attojoule}. An ongoing discussion \cite{EITfavor,Steinberg} about the impossibility of attaining larger XPM using this EIT mechanism began about eight years ago. A description of the XPM phenomenon considering multimode beams and the response time of the medium implies that the noise on the phase shift, due to the photon number and phase complementarity, would compromise the fidelity of the operation \cite{Shapiro1,Shapiro2,Shapiro3}. It was also suggested that in order to strengthen the nonlinear response a smaller EIT bandwidth is required, which also creates a slow-medium condition increasing the response time \cite{Gea-Banacloche}. Mismatch of the group velocity between the control and the probe pulses has been proven to be a undesired issue as well \cite{Shapiro2014}. %Averaging is fine, single pgotons shots might not~\cite{PhotonCounting}.   

Despite the controversy, XPM still remains a good candidate for an all-optical deterministic logic gate. For example, different cavity-based protocols have been proposed \cite{Purcell2007,Yang2014}, and a cavity-EIT system using an ensemble of laser-cooled atoms has been recently used to create an equal-time cross-modulation between two weak beams \cite{Beck2014,KristiXPM}. A different approach using a high optical depth ($\sim$100) inside a hollow-core photonic bandgap fibre filled with Rb atoms was also proposed \cite{Gaeta}. Using large detuning ($\sim$700 MHz) from the single-photon resonance in a ladder system, a cross-phase shift of 0.3 mrad per average photon with a response time of $\sim$5 ns was reported. 

Besides XPM, the possibility of nonlinear quantum control spans a broad spectrum of technological applications, where more elaborate designs based on cavity QED, Electromagnetically Induced Transparency (EIT) or a mixture of both (cavity-EIT) have been proposed, and also with the possibility of using atomic Rydberg states or nanophotonic systems. A review of these efforts can be found in Ref. \cite{LukinReview}.

We introduce in this letter an alternative and still unexplored approach to XPM based on a classical description of the field interacting with an atomic ensemble inside an optical resonator. %We start with the well known semi-classical expression for the cross susceptibility, and do not require the use of EIT. 
We set the conditions such that the population of the atomic ensemble remains in the ground state so the known cross-Kerr nonlinear electric susceptibility for a ladder system can be used. We will show that this approach falls in the interesting regime of a far-off-resonance effective cooperativity of unity ($\eta_{eff}=1$) independently of the value of the on-resonance cooperativity. This protocol avoids most of the problematic issues of previous approaches mentioned above since (i) it does not require EIT, (ii) only one transverse cavity mode of the control and probe beams interacts with the atomic ensemble, (iii) the control beam acts as a switch for the phase-shift on the probe beam which acquires no self-phase modulation since the electronic population is on the ground state (see Fig. \ref{fig:Setup}a), and (iv) no elaborated experimental set-ups are required. In addition, even though the protocol is presented for coherent continuous waves, (v) the system is operated very far-off-resonance so no mismatch of group velocities for the control and the probe beams is induced, and fast response time is expected so extension to using short pulses (with cavity-lifetime bandwidth) could be eventually implemented. This proposal satisfies the conditions necessary to realize optical quantum computation using weak nonlinearities as proposed by Munro, et al. \cite{Munro2005}, non-local interferometry as proposed by Kirby and Franson \cite{Kirby2013,Kirby2014}, and device-independent quantum key distribution~\cite{CurtisQKD}. It also supports some recent experimental efforts \cite{Hickman,Kirby2015,XPM-UMBC}.      

This letter is organized as follows: In section \ref{non-cavity-case} we review a complete analysis of the one-pass XPM as a function of the on-resonance optical depth. We show that large detuning allows for considerable phase shifts with better transmission of the control beam when compared to the close-to-resonance condition, which is more commonly used. In section \ref{cavity-case} we evaluate the advantages of coupling the system to an optical resonator and show that large XPM with controllable transmission is possible. In addition, we show how the semi-classical description is self-consistent with a far-off-resonance effective cooperativity equal to one. Finally we present our conclusions.  

\section{Single-Pass Cross-Phase Modulation (Review)}\label{non-cavity-case}
\begin{figure}[htbp]
\centering
%\fbox{\includegraphics[width=\linewidth]{Setup}}
\includegraphics[width=\linewidth]{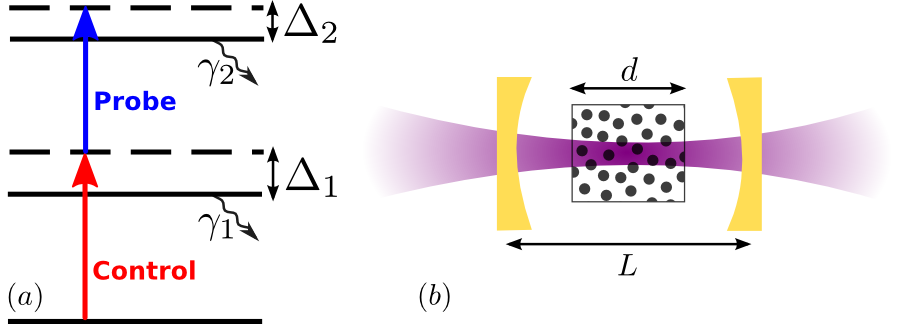}
\caption{Schematic of the ladder-based XPM setup. One weak (few-photon level) control laser beam induces a phase-shift in a probe laser beam, mediated by an ensemble of atoms inside an optical resonator of length $L$. $\Delta_1$ is the single-photon detuning and $\Delta_2$ is the two-photon detuning. Description without (with) an optical resonator is given in section \ref{non-cavity-case} (\ref{cavity-case}).}
\label{fig:Setup}
\end{figure}

In order to evaluate the advantage of using an optical resonator, we first study the free-space (non-cavity) scenario. We estimate the phase induced in a probe laser beam by a weak (few-photon level) control coherent beam in an atomic ladder system, as shown in Fig. \ref{fig:Setup}a. Two Gaussian TEM$_{00}$-mode beams interact with an atomic ensemble through a distance $d$ (non-cavity case in Fig. \ref{fig:Setup}b), so the Non-Cavity phase modulation is \cite{Boyd2008207}
\begin{equation}
\phi^{NC}=k_pn_2\int_{0}^{d}I_c(z)dz,
\end{equation}
where $k_p$ is the probe wave-number, $n_2=3\mbox{Re}\left(\chi^{(3)}\right)/2n_0^2\epsilon_0c$ is the second-order nonlinear cross-refractive index, and $I_c(z)$ is the intensity of the control beam. We assume a Doppler-free configuration for the linear and non-linear susceptibilities, and only the control beam suffers from linear absorption. Two-photon absorption is not introduced since low intensities for both beams are considered. The expression for the third-order susceptibility is given by \cite{shenbook,Gaeta}
\begin{equation}
\chi^{(3)}=-\frac{N\mu_1^2\mu_2^2}{\epsilon_0\hbar^3\gamma_1^2\gamma_2}\frac{1}{\left(i+\delta_1\right)^2\,\left(i+\delta_2\right)},
\end{equation}
where $N$ is the (atomic) number density, $\mu_1$ and $\mu_2$ are the dipole moments of both transitions, and $\gamma_1$ and $\gamma_2$ are the transitions' population decay rates. To make the calculations general to any atomic system choice, we work using the dimensionless \textbf{\textit{relative single-photon detuning}} $\delta_1=\Delta_1/\gamma_1$ and the \textbf{\textit{relative two-photon detuning}} $\delta_2=\Delta_2/\gamma_2$, and find their optimal values for which the cross phase (XP) is a maximum. 

We consider both beams roughly collimated and an Optical Depth for the control beam given by $OD_{NC}(\delta_1)=OD/\left(1+\delta_1^2\right)$, i.e. $I_p(z)=I_p$ and $I_c(z)\approx I_0\,e^{-OD_{NC}(\delta_1)z/d}$ during the distance $d$. Here 
\begin{equation}
OD=\frac{k_c N\mu_1^2 d}{\hbar\epsilon_0\gamma_1}=\frac{3Nd\lambda_c^2}{2\pi}
\end{equation}
is the \textit{\textbf{on-resonance optical depth}}. We have assumed the interaction length $d$ to be smaller than the Rayleigh length of both beams. The XP for the \textit{Non-Cavity} case takes the form
\begin{equation}\label{eq:PhaseNC}
\phi^{NC}(\delta_1,\delta_2;OD)=\phi_{max}\,f(\delta_1,\delta_2;OD),
\end{equation}
where
\begin{equation}\label{functionf}
f(\delta_1,\delta_2;OD)=\left[1-e^{-OD/(1+\delta_1^2)}\right]\frac{-\delta_1^2\delta_2+2\delta_1+\delta_2}{\left(1+\delta_1^2\right)\left(1+\delta_2^2\right)},
\end{equation}
such that $|f(\delta_1,\delta_2;OD)|\leq1$, and
\begin{equation}\label{phimax}
\phi_{max}=\frac{3\mu_2^2}{2\epsilon_0 c\hbar^2\gamma_1\gamma_2}\left(\frac{\lambda_c}{\lambda_p}\right)I_0.
\end{equation}
$I_0$ is the input intensity of the collimated control beam and $\lambda_p$ ($\lambda_c$) is the wavelength of probe (control) beam. Note that eq. (\ref{eq:PhaseNC}) is independent of the probe's intensity $I_p$ and is only valid when below the atomic saturation limit.  

$\phi_{max}$ is the maximum possible total cross phase induced in the probe beam by one pass of the control beam and it depends explicitly upon the choice of the atomic ladder transitions ($\mu_2,\gamma_1,\gamma_2$), the respective lasers' wavelengths ($\lambda_c,\lambda_p$) and the control beam input intensity ($I_0$). The average XP per single control photon per atomic cross section is usually a good measure of the strength of the interaction. This value can be obtained in our case by setting $I_0=2\pi \hbar c\gamma_2/\lambda_c\sigma_c$ in eq. (\ref{phimax}), where we have assumed that the response time of the XPM is set by the two-photon relaxation time, and $\sigma_c=3\lambda_c^2/2\pi$ is the atomic cross section of the control beam. The average cross phase per photon per cross section takes the form, 
\begin{equation}
\bar{\phi}_{pp}=\frac{2\pi^2\mu_2^2}{\epsilon_0\hbar\gamma_1\lambda_p\lambda_c^2}.
\end{equation}

For example, for the $5S_{1/2}\rightarrow5P_{3/2}\rightarrow5D_{5/2}$ transition in Rb used in Ref.~\cite{Gaeta}, $\phi_{max}/I_0\approx\,23\,rad/(nW/\mu m^2)$ and $\bar{\phi}_{pp}\approx88\,mrad$, where $\mu_2\approx8.4\times 10^{-30}$ C$\cdot$m, $\gamma_1\approx2\pi(6\,MHz)$, $\gamma_2\approx2\pi(0.67\,MHz)$, $\lambda_c\approx780.2$ nm and $\lambda_p\approx776$ nm. Such a numerical value for $\phi_{max}$ means that a 1 $nW$ control laser beam with a cross section of 100 $\mu m^2$ would induce a maximum XP in the probe beam of 0.23 $rad$, or equivalently 256 $\mu rad$ per averaged photon. Note that $\phi_{max}$ is \textit{independent of the Optical Depth} showing the necessity of using techniques like EIT for example, where the $OD$ dependence emerges. We will introduce in the next section a different approach, where XPM enhancement is due to the many intracavity bounces of the fields. 

\begin{figure}[htbp]
\centering
\includegraphics[width=\linewidth]{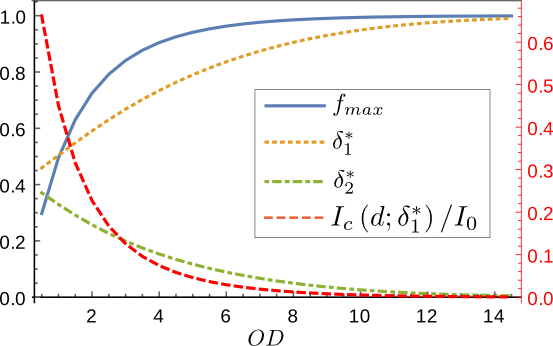}
\caption{Maximization of the XPM for the one-pass system as a function of the on-resonance optical depth. The left (black) scaling is for $f_{max}$, $\delta_1^*$ and $\delta_2^*$, and the right (red) scaling is for the transmission of the control beam.}
\label{fig:MaximumNC}
\end{figure}

The phase $\phi_{max}$ defines the maximum possible value for $\phi^{NC}$, and the function $f$ carries the dependence on $\delta_1$, $\delta_2$ and $OD$. We first find the parameters for which $f(\delta_1,\delta_2;OD)$ is a maximum ($f(\delta_1^*,\delta_2^*;OD)=f_{max}$) and give the optimal XPM. It is important to note that $f(-\delta_1,-\delta_2;OD)=-f(\delta_1,\delta_2;OD)$, so for every couple ($\delta_1^*,\delta_2^*$) such that $\phi^{NC}$ is maximum, ($-\delta_1^*,-\delta_2^*$) gives the same optimal phase but with opposite sign. We only refer to the positive pair of coordinates during the rest of this section since the symmetry is clear. Fig. \ref{fig:MaximumNC} shows the maximization of $f(\delta_1,\delta_2;OD)$ as a function of OD. The maximum value of $f$ grows as $OD$ increases converging to unity when $OD\gtrsim 10$, and the values for $\delta_1^*$ and $\delta_2^*$ converge to $1$ and $0$ respectively in a slower fashion. This means that in order to reach the maximum phase $\phi_{max}$ it is sufficient to have an on-resonance optical depth equal or larger than $\sim$15 and tune the lasers' frequencies to exactly the two-photon resonance ($\delta_2=0$) but one linewidth detuned from the single-photon transition ($\delta_1=1$ or $\Delta_1=\gamma_1$). To avoid considerable two-photon absorption during an actual experiment it might be expected to operate off-resonance from the two-photon transition. The system is however inefficient since the control beam is highly absorbed (red-dashed line in Fig. \ref{fig:MaximumNC} is practically null for $OD\geq15$) close to resonance. Note that a smaller value of OD could be used to obtain better transmission. For example, for $OD=1$ the transmission can be made up to $45\%$, paying the price with a lower XP of $\sim0.5\phi_{max}$.  
%This type of Zeno gate has been studied before, for example in Ref. \cite{You2012}. We do NOT evaluate here how this process could be improved by the use of Electromagnetically Induced Transparency (EIT) in order to overcome such absorption of the Signal, nor we evaluate the speed at which XPM is happening. 

\begin{figure}[htbp]
\centering
\includegraphics[width=\linewidth]{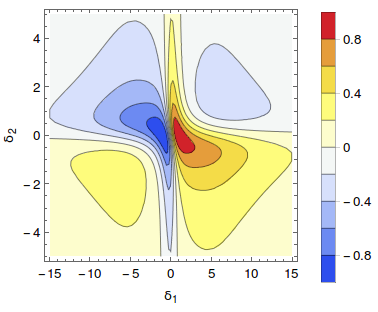}
\caption{Contour plot of $f(\delta_1,\delta_2;100)$. Global maximum and minimum are observed very close to the origin, but local and broader peaks emerge in the diagonal direction.}
\label{fig:ContourMinimumNC}
\end{figure}

\begin{figure}[htbp]
\centering
\includegraphics[width=\linewidth]{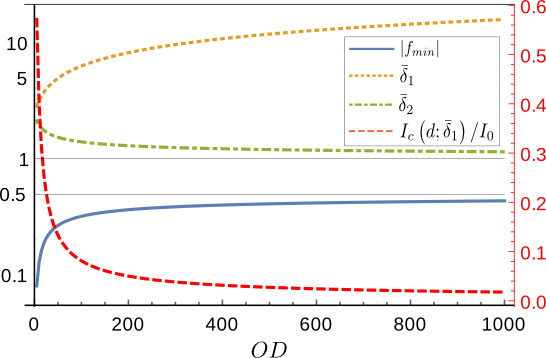}
\caption{Local minimum of $f(\delta_1,\delta_2;OD)$ as a function of $OD$, where $f_{min}=f(\bar{\delta}_1,\bar{\delta}_2;OD)$. The left (black) logarithmic scaling is for $|f_{min}|$, $\overline{\delta}_1$ and $\overline{\delta}_2$, and the right (red) scaling is for the transmission of the control beam. $|f_{min}|$ converges to 0.5 and $\bar{\delta}_2$ to 1 respectively.}
\label{fig:MinimumNC}
\end{figure}

An interesting feature of eq. (\ref{functionf}) is that for large $OD$ values a local (and much broader) minimum of $f$ starts to appear for large values of $\delta_1$. For example, Fig. \ref{fig:ContourMinimumNC} shows a contour plot of $f$ for $OD=100$. The maximum of $f$ is located at $(\delta_1^*,\delta_2^*)=(1,0)$ as shown before (Fig. \ref{fig:MaximumNC}), but a local minimum with value $f_{min}=f\left(\bar{\delta}_1,\bar{\delta}_2;OD\right)\approx-0.32$ raises when $\bar{\delta}_1\approx6.23$ and $\bar{\delta}_2\approx1.38$, indicating that $\phi^{NC}=-0.32\phi_{max}$ if $\Delta_1\approx6.23\gamma_1$, $\Delta_2\approx1.38\gamma_2$ and $OD=100$. The results are easily extended as a function of $OD$  and are shown in Fig. \ref{fig:MinimumNC}. $f_{min}$ converges to $-0.5$ and $\overline{\delta}_2$ to $1$ as $OD$ grows, nevertheless this happens quite slow after $OD\sim100$. For example, if $OD=10000$ we find that $f_{min}\approx-0.47$, $\overline{\delta}_2\approx1.05$ and $\overline{\delta}_1\approx43$. In other words, extremely large values of $OD$ are required to obtain a XPM of only $\phi^{NC}=-0.5\phi_{max}$. In addition, the sum of both laser's frequencies must be slightly off resonance of the two-photon transition ($\delta_2\rightarrow1$ or $\Delta_2\rightarrow\gamma_2$) and a large single-photon detuning ($\bar{\delta}_1$) is require to avoid large absorption of the control beam (red-dashed line in Fig. \ref{fig:MinimumNC}).%, although such absorption is not as strong as in the maximization case (Fig. \ref{fig:MaximumNC}). 

This minimization configuration is reasonably interesting~\cite{You2012} since it is performed far off resonance. Nevertheless, it does not offer better XP than the near-resonance case. %Nevertheless it requires very large $OD$ and offers no actual gain in the obtained cross phase. 
A similar far-off-resonance configuration was used in Ref. \citep{Gaeta} for a Doppler-broadened linear absorption of the control beam in a hollow-core photonic band-gap fibre filled with Rb atoms. The reported optimal result was  $\phi/I_0\sim 0.14\,rad/(nW/\mu m^2)$ when $\delta_1=\delta_2\approx14$. We note that our theory predicts a local maximum of $\phi_{max}/2I_0\sim 0.02\,rad/(nW/\mu m^2)$ for the given experimental parameters. Such a discrepancy is possible since we use a Lorentzian Doppler-free linear absorption, and ignore two-photon absorption. The short atomic transient time governing the photon emission sets the upper value for the XP in the hollow-core photonic bang-gap fibre configuration, nevertheless the advantage relies on the high allowed values for the control-beam intensity $I_0$.   

\section{Cavity or Multi-Pass Cross-Phase Modulation}\label{cavity-case}
In the previous section we saw how a local peak for XPM can be induced very far off resonance with better transmission of the control beam than the near-resonance situation. This condition requires large OD values, which are rare in laser-cooled atomic ensembles. We show now a considerable increase in the XPM if the atom ensemble interacts with the fields inside a doubly resonant optical cavity. For this case, the far-off-resonance condition will emerge when avoiding large absorption due to the many bounces of the field inside the resonator.  

\begin{figure}[htbp]
\centering
\includegraphics[width=\linewidth]{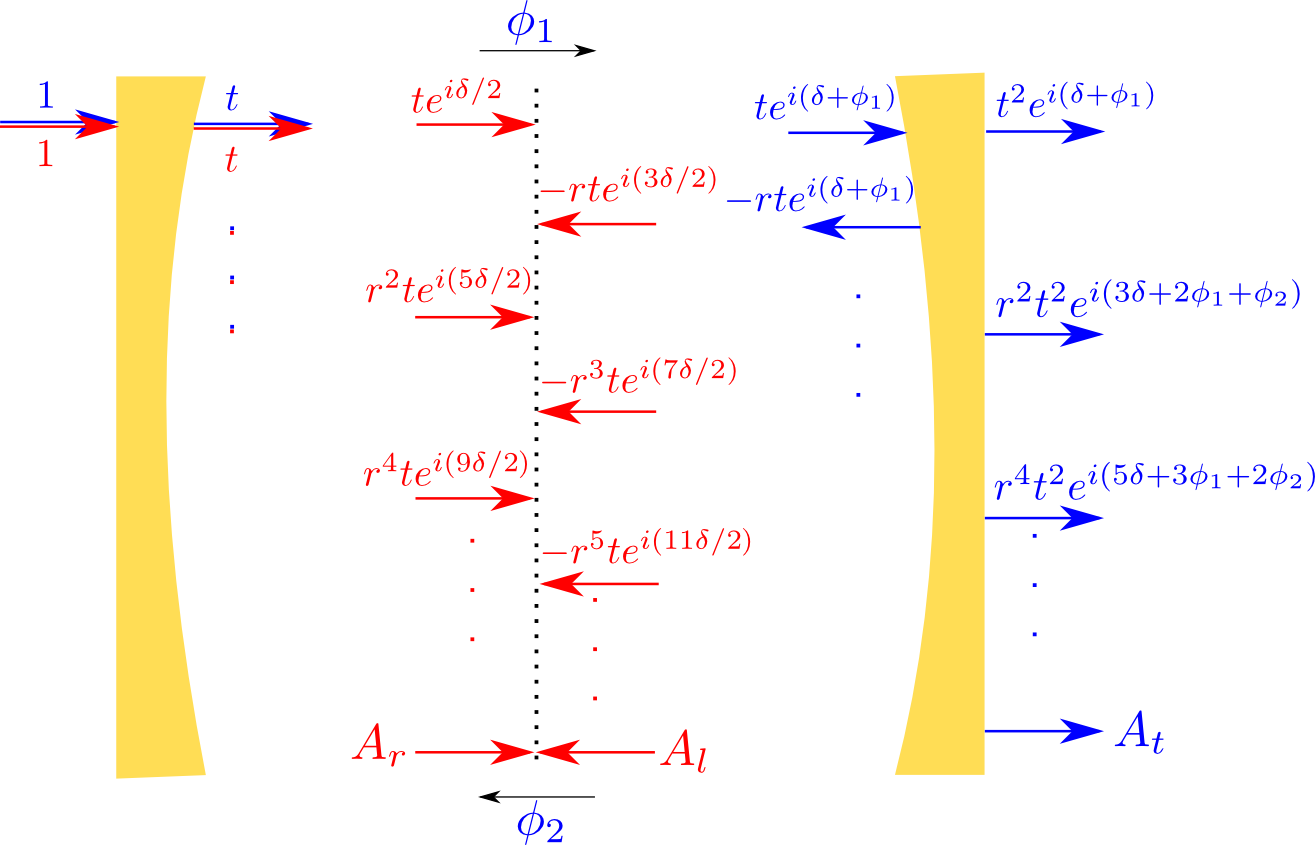}
\caption{Cartoon of the intracavity fields using simple ray optics. Control (red) and probe (blue) beams, with normalized input amplitudes, enter the resonator through one of the mirrors with reflectivity $R=|r|^2=1-|t|^2$. Only the ray components necessary for the calculations are shown.}
\label{fig:Cavity}
\end{figure}

We consider the system depicted in Fig. \ref{fig:Cavity}. Both beams, control and probe, enter from the left a cavity designed with identical mirrors of reflectivity $R=|r|^2=1-|t|^2$ and interact with the atomic ensemble at the center of the resonator. For the probe beam, we assume it is not absorbed and that it acquires a phase shift every time it passes trough the center due to cross-Kerr interaction with the atoms and the control beam. In general, the induced phase when it propagates to the right ($\phi_1$) is different than the one when it propagates to the left ($\phi_2$), defining the output probe field as   
\begin{eqnarray}
A_t&=&t^2e^{i(\delta+\phi_1)}\left[1+r^2e^{i(2\delta+\phi_1+\phi_2)}+r^4e^{i2(2\delta+\phi_1+\phi_2)}+...\right]\nonumber\\
&=&\frac{(1-R)e^{i(\delta+\phi_1)}}{1-Re^{i(2\delta+\phi_1+\phi_2)}}\label{TransmissionAmplitde},
\end{eqnarray}
where $\delta=k_pL$ is the propagation phase due to every pass and $L$ is the resonator's length. Then, the total global phase of the probe beam at the output of the resonator is given by   
$-i\ln\left(A_t/|A_t|\right)=\delta+\phi_1+(i/2)\ln{\left[\left(1-R\,e^{i(2\delta+\phi_1+\phi_2)}\right)/\left(1-R\,e^{-i(2\delta+\phi_1+\phi_2)}\right)\right]}$. The induced cavity-case nonlinear phase is defined when $\delta=0$,
\begin{equation}
\phi^C=\phi_1+\frac{i}{2}\ln\left[\frac{1-R\,e^{i(\phi_1+\phi_2)}}{1-R\,e^{-i(\phi_1+\phi_2)}}\right]\approx\frac{\phi_1+R\,\phi_2}{1-R},
\end{equation} 	
since $\{\phi_1,\phi_2\}\ll1$. This result is intuitive since $(1-R)^{-1}$ is the average number of passes trough the center that a photon does inside the resonator before escaping. Also, after a photon acquires a phase $\phi_1$ it is reflected two more times acquiring the phase $\phi_2$ before it acquires $\phi_1$ again. 

The phases $\phi_1$ and $\phi_2$ are determined by the intracavity fields $A_l$ and $A_r$, and by the one-pass induced cross phase of Eq.~(\ref{eq:PhaseNC}),
\begin{eqnarray}\label{approx}
\phi_1&=&|A_l|^2\,\phi^{NC},\\
\phi_2&=&|A_r|^2\,\phi^{NC},
\end{eqnarray}
where $A_l$ is the total control field (normalized to the input) that propagates to the left inside the resonator right before passing through the center of it, and $A_r$ is the equivalent component propagating to the right. Since we are interested in the maximum of these intracavity intensities we ignore the self- and cross-phase modulation of the control beam. Nevertheless, an imaginary term is introduced in the propagation phase, $\delta\rightarrow k_cL+iOD_{NC}(\delta_1)/2$ (with $k_c=2\pi/\lambda_c$), to account for absorption after every pass through the atomic gas,
\begin{eqnarray}
|A_r|^2&=&\frac{(1-R)\,e^{-OD_{NC}(\delta_1)/2}}{\left[1-Re^{-OD_{NC}(\delta_1)}\right]^2+4Re^{-OD_{NC}(\delta_1)}\sin^2(k_cL)},\nonumber\\
|A_l|^2&=&R\,e^{-OD_{NC}(\delta_1)}\,|A_r|^2.\nonumber
%\frac{R(1-R)\,e^{-3\,OD_{NC}(\delta_1)/2}}{\left[1-Re^{-OD_{NC}(\delta_1)}\right]^2+4Re^{-OD_{NC}(\delta_1)}\sin^2(k_cL)}.
\end{eqnarray}

The optimal cross phase is obtained for the maximum possible values of $|A_{\{r,l\}}|^2$, when $k_cL=\{0,\pi,2\pi,...\}$, 
\begin{eqnarray}\label{factorchi}
\phi^C&=&\frac{|A_l|^2+R|A_r|^2}{1-R}\,\phi^{NC}\nonumber\\
%&=&\frac{R}{1-R}\left[1+e^{-OD_{NC}(\delta_1)}\right]|A_r|^2\,\phi^{NC}\nonumber \\
&=&R\,\frac{e^{-OD_{NC}(\delta_1)/2}\left[1+e^{-OD_{NC}(\delta_1)}\right]}{\left[1-Re^{-OD_{NC}(\delta_1)}\right]^2}\,\phi^{NC}.
\end{eqnarray}
Using eq. (\ref{factorchi}) we define an effective non-linear electric susceptibility due to coupling to the optical resonator as
\begin{equation}
\mbox{Re}\left(\chi^{(3)}_{eff}\right)=\alpha(x,R)\,\mbox{Re}\left(\chi^{(3)}\right),
\end{equation}
where we call $x=OD_{NC}=OD/(1+\delta_1^2)$ for notation simplicity, and $\alpha(x,R)=R\,e^{-x/2}(1+e^{-x})/(1-Re^{-x})^2$ is the factor in front of $\phi^{NC}$in Eq.~(\ref{factorchi}). The one-pass optical depth at a given detuning, $x$, plays the role of a control parameter. For example, if $x\gg1$, $\alpha\approx Re^{-x/2}$, which is a undesirable case since absorption gets amplified by the many bounces inside the resonator. On the other hand, small single-pass absorption, $x\ll1$, amplifies the non-linear response by a factor of $\alpha\approx2F^2/\pi^2$, where $F\approx\pi\sqrt{R}/(1-R)$ is the \textbf{\textit{finesse of the resonator}}. This dependence on the finesse squared is the basis of our protocol to find a maximization of the XPM for small values of $x$ (small one-pass absorption). 

We have assumed in this section that the atomic intermediate state lifetime is larger than the cavity lifetime, i.e. $\gamma_1<c/(2LF)$. This assumption means that one either needs to use a long life-time atomic state (small $\gamma_1$), or to limit the cavity length and finesse such that $LF<c/(2\gamma_1)$.

The cavity-enhanced XP of eq. (\ref{factorchi}) can be expressed as
\begin{equation}
\phi^C(\delta_1,\delta_2;OD,R) =\phi_{max}\;g\left(\delta_1,\delta_2;OD,R\right),
\end{equation}
where $\phi_{max}$ is the maximum XP for the Non-Cavity case as in eq. (\ref{phimax}), and
%\begin{strip}
\begin{widetext}
\begin{equation}
g\left(\delta_1,\delta_2;OD,R\right)=-R\,e^{-OD_{NC}(\delta_1)/2}\,\frac{1-e^{-2\,OD_{NC}(\delta_1)}}{\left[1-R\,e^{-OD_{NC}(\delta_1)}\right]^2}\,\frac{\delta_1^2\delta_2-2\delta_1-\delta_2}{(1+\delta_1^2)(1+\delta_2^2)}
\end{equation}
\end{widetext}
%\end{strip}
allows a direct comparison of the performance versus the one-pass system (section \ref{non-cavity-case}). In contrast to the function $f$, the magnitude of $g$ is not upper bounded. A proper maximization of the cross phase $\phi^C$ is now introduced where the optimal detuning is defined for a given value of $OD$.        

\subsection{Maximization of XPM in a Cavity}
\begin{figure}[htbp]
\centering
\includegraphics[width=\linewidth]{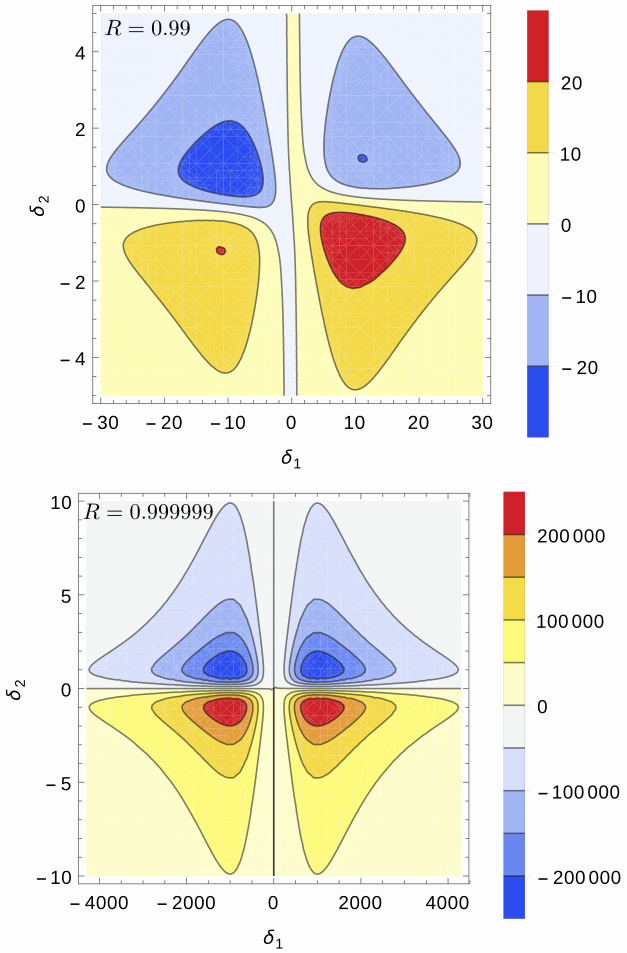}
\caption{Contour plots of $g(\delta_1,\delta_2;1,R)$ for $R=0.99$ (upper) and $R=0.999999$ (lower). Global (local) maximum and minimum are seen in the antidiagonal (diagonal) axis. All peaks are located close to $\delta_2=\pm 1$ independently of $R$, but the absolute values of $g$ at the peaks and $\delta_1^*$ grow with $R$.}
\label{fig:ContourMaximumC}
\end{figure}

For the Non-Cavity case, we studied the function $f(\delta_1,\delta_2;OD)$. We focus now on the maximization of $g(\delta_1,\delta_2;OD,R)$. For $OD=1$ for example, Fig. \ref{fig:ContourMaximumC} shows contour plots of  $g(\delta_1,\delta_2;1,R)$ for $R=0.99$ ($F=313$) and $R=0.999999$ ($F=3.14\times10^6$). The first thing to note here is that the location of the global and local peaks is symmetric respect to the detuning $\delta_1$ and that the sign of the phase shift is determined by the sign of $\delta_2$. Also, the required detuning $\delta_1^*$ for the optimal XPM and the peak values increase proportional to the finesse of the cavity.

In fact, if $R\approx1$ (very high finesse cavity) the plot has a nearly symmetric shape and all four peaks have approximately the same absolute values, located at coordinates $\left(\delta_1^*,\delta_2^*\right)\approx\left(\pm\sqrt{OD\cdot F/\pi},-1\right)$ and $\left(\bar{\delta}_1,\bar{\delta}_2\right)\approx\left(\pm\sqrt{OD\cdot F/\pi},1\right)$. Here we have assumed that $\delta_1^2\gg1$, meaning a very large single-photon detuning, $\Delta_1\gg\gamma_1$. For simplicity, we focus from now on only on the global maximum, where $\delta_1^*\gg1$ and $\delta_2^*\approx-1$. This allows us to calculate the maximum possible value of $g$ since $\exp{\left[-OD_{NC}(\delta_1^*)\right]}\approx R$,
%\begin{eqnarray}
%g_{max}&=&g\left(\delta_1^*,-1;OD,R\right)\nonumber\\
%&\approx&-R\sqrt{R}\frac{1-R^2}{(1-R^2)^2}\left[\frac{-(\delta_1^*)^2-2\delta_1^*+1}{2\left[1+(\delta_1^*)^2\right]}\right]\nonumber\\
%&\approx&\frac{R\sqrt{R}}{1-R^2}\left(\frac{(\delta_1^*)^2}{2(\delta_1^*)^2}\right)\nonumber\\
%&=&\frac{R\sqrt{R}}{2(1+R)(1-R)}\nonumber\\
%&\approx&\frac{F}{4\pi}\label{amplification}
%\end{eqnarray}
\begin{equation}
g_{max}= g\left(\delta_1^*,-1;OD,R\right)\approx\frac{F}{4\pi},\label{amplification}
\end{equation}
where we assume that $OD\ll(\delta_1^*)^2\approx\left(F/\pi\right)OD\gg1$, showing that the value of the $OD$ is irrelevant for the optimization as long as a high finesse resonator is used. Eq. (\ref{amplification}) is the major result of this letter and shows that an optical resonator increases the maximum possible XPM by a factor of $F/4\pi$ with respect to the one-pass non-cavity case. Note that the maximum XPM in the cavity system takes the form 
\begin{equation}
\phi^{C}_{max}\approx\frac{F}{4\pi}\,\phi_{max},
\end{equation}
and it is independent of the $OD$. %Fig. \ref{fig:Gain} shows this behaviour and also the OD-dependent detuning $\delta_1^*$. 
The required detuning is given by $(\delta_1^*,\delta_2^*)\approx(\sqrt{OD\cdot F/\pi},-1)$. For our example, as in section~\ref{non-cavity-case}, in the $5S_{1/2}\rightarrow5P_{3/2}\rightarrow5D_{5/2}$ two-photon transition in Rb we get that $\phi_{max}^{C}/I_0\approx (1.8\times F)\,rad/(nW/\mu m^2)$ and $\bar{\phi}_{pp}^C\approx(7\times F)\,mrad$. This result for Rb means that if a 1 $nW$ control beam with cross section of $100\,\mu m^2$ is used, a cavity with a finesse of only $\sim$175 would be enough for inducing a $\pi$ phase shift in the probe beam, or equivalently 3.6 mrad per averaged photon.
%\begin{figure}
%\centering
%\includegraphics[scale=0.6]{Gain.png}
%\caption{Cavity-enhancement, $g_{max}=F/4\pi$, of the maximum XPM respect to the one-pass case, and the required detuning $\delta_1^*=\sqrt{OD\cdot F/\pi}$ of the intermediate atomic level. Both as a function of the finesse and the reflectance of the mirrors of the resonator.}
%\label{fig:Gain}
%\end{figure}
\subsection{Absorption of Control Beam}
We are now interested on the transmission through the resonator of the control beam. Using eq. (\ref{TransmissionAmplitde}) for $\phi_1=\phi_2=0$ and modifying it as $\delta\rightarrow k_cL+iOD_{NC}(\delta_1)/2$, we define the transmission of the control beam as 
\begin{equation}
T_c=\frac{(1-R)^2e^{-OD_{NC}(\delta_1)}}{(1-Re^{-OD_{NC}(\delta_1)})^2+4Re^{-OD_{NC}(\delta_1)}\sin^2(k_cL)}.
\end{equation}

Note that if no absorption is assumed $(OD=0)$ the well-known result is recovered, $T_c\rightarrow [1+(2F/\pi)^2\sin^2(k_cL)]^{-1}$~\cite{salehbook}. The maximum transmission is then given by
\begin{equation}
T_c^{max}=\left(\frac{1-R}{1-R\,e^{-OD_{NC}(\delta_1)}}\right)^2e^{-OD_{NC}(\delta_1)},
\end{equation}
which allows us to define the cavity-effective optical depth,
\begin{eqnarray}
OD_C(\delta_1)&=&-\ln\left(T^{max}_c\right)\nonumber\\&=&OD_{NC}(\delta_1)+2\ln\left[\frac{1-Re^{-OD_{NC}(\delta_1)}}{1-R}\right].
\end{eqnarray}
For the optimized cavity-XPM this optical depth takes the form $OD_{C}(\delta_1^*)\approx -\ln R+\ln 4\approx \ln4$, which gives a control beam's transmission of $\exp{\left[-OD_C(\delta_1^*)\right]}\approx0.25$ (or 25$\%$) for $R\approx1$.

%As high finesse resonators are required for the enhancement process, we note that the maximum XPM is obtained for a transmission of the control field of only $\sim$25$\%$. 
We now evaluate how the XPM becomes compromised when higher transmission is required. We desire a control beam's transmission given by $T_c(x,R)=1-\epsilon$, with $\epsilon\ll1$ and $T_c(x,R)=e^{-x}\left[(1-R)/\left(1-R\,e^{-x}\right)\right]^2$. For our case of interest of large detuning, $x\ll1$, the transmission can be expressed as $T_c(x,R)\approx 1-(1+R)x/(1-R)$, so $x=(1-R)\,\epsilon/(1+R)\approx(1-R)\,\epsilon/2$. Then if $\delta_2=-1$ we can approximate $g(x,R)\approx x/(1-R)^2$ for $x\ll1$, giving an amplification by a factor of $g_\epsilon= \epsilon(F/2\pi)=2\epsilon g_{max}$ in the XP, and $\delta_1^{(\epsilon)}=\sqrt{(2/\epsilon)(F/\pi)OD}=\sqrt{2/\epsilon}\,\delta_1^*$. For example, if we want to boost the transmission of the control beam from 25$\%$ to $90\%$, i.e. $\epsilon=0.1$, the XP gets amplified by a factor of $F/20\pi$ respect to the optimal regime in the non-cavity case, and it is only five times smaller than in the optimal cavity regime. This shows that transmission can be made very close to unity, but very high finesse for the resonator is required to still obtain a considerable XPM.
 
\subsection{Effective Cooperativity}

Cavity Quantum Electrodynamics (cavity QED), which describes a system composed of one two-level atom coupled to a single mode of an optical resonator, has been the building block for a vast variety of approaches in the ongoing goal of attaining full quantum control at the single atom and photon level. The primary requirement of such approaches is to satisfy the atom-photon strong-coupling regime at the single-photon level \cite{Kimble1998,Kimble2005}. This regime is defined by a large on-resonance (single-atom) cooperativity value, namely $\eta>1$. A cooperativity exceeding unity is normally understood to be when quantum phenomena play an important role in the system's dynamics \cite{RabiSplitting}. Strong photon-photon nonlinearities can be obtained because one photon is able to saturate the atomic response, and coherent control overcomes photon leaking out of the resonator and spontaneous emission \cite{PhotonBlockade}. Nevertheless, it was recently shown that many effects in multi-atoms cavity QED can be understood from a fully classical description, even within the strong-coupling regime \cite{ClassicalCavity}.

In cavity QED the on-resonance single-atom cooperativity of a two-level atom and one photon is defined as $\eta=2g_0^2/\kappa\gamma$, where $2g_0=2\mu\sqrt{\omega/2\hbar\epsilon_0 V_m}$ is the dipole coupling rate or single-photon Rabi frequency, $2\kappa=\pi c/LF$ is the cavity-field damping rate or bandwidth of the resonator, $\gamma=\omega^3\mu^2/6\pi\epsilon_0\hbar c^3$ is the transverse incoherent atomic decay rate to non-cavity modes, and $V_m=A_mL$ is the cavity mode volume. Values of cooperativity larger than one means that coherent Rabi oscillations dominate over decoupling due to spontaneous emission and over photon-leaking out of the cavity through one of the mirrors. 

Alternatively, the cooperativity can be written as $\eta=(4F/\pi)(\sigma/2A_m)$, where $\sigma=3\lambda^2/2\pi$ is the atomic cross section. Two important aspects arise from this definition: (i) the free-space cooperativity $\eta_{fs}=\sigma/2A_m$ is usually a very small quantity and can be understood as the probability for a photon to be scattered by one atom, and (ii) by coupling the atom-photon system to a resonator this free-space cooperativity gets enhanced by $4F/\pi$, the same amount that the intracavity field intensity is amplified with respect to the input intensity.        

An effective way to increase the probability that a photon gets scattered is increasing the number of atoms inside the resonator, and it was recently shown \cite{ClassicalCavity} that most interaction processes can be understood as an effective cooperative enhancement by half the average number of atoms inside the cavity mode. The term \textquotedblleft effective\textquotedblright  here comes from the fact that the photon is very likely  to get scattered by the many atoms ensemble, but this does not change the linear response of the medium as it happens in the strong-coupling cavity QED regime. For our purposes of giving a classical description, we define the \textit{\textbf{far-detuned effective cooperativity}} for the control beam as
\begin{equation}
\eta_{eff}(\delta_1)=\left(\frac{NA_md}{2}\right)\frac{\eta}{1+\delta_1^2}=\left(\frac{F}{\pi}\right)\frac{OD}{1+\delta_1^2},
\end{equation}
which takes the interesting form of the one-pass optical depth multiplied by the average number of passes of a photon through the atomic ensemble inside the resonator, $F/\pi$.

The boundary $\eta=1$ normally marks the transition from a bad to a strong cavity regime in cavity QED. So, we find it interesting that the effective cooperativity $\eta_{eff}$ takes the value of one when the XPM is maximum, $\delta_1=\delta_1^*\approx\sqrt{(F/\pi)OD}\gg1$. However, $\eta_{eff}=1$ is not equivalent to the strong coupling regime in cavity QED. In fact, the value of $\eta_{eff}=1$ is always obtained independently of the value of $\eta$, which must be smaller than one for our protocol to remain valid. Note that the on-resonance cooperativity, $\eta$, is defined as a purely geometrical factor which depends on the reflectivity of the resonator's mirrors and on how tightly a beam is focused. Our introduced effective cooperativity depends also upon how strong the photon-scattering process is, which is based on the number of atoms interacting with the laser mode and on the detuning from the atomic transition. Importantly, the value of $\eta_{eff}=1$ for the optimal XPM emerges independently of the value of the on-resonance optical depth $OD$ and of the finesse of the resonator $F$. 

\section{Conclusions}

We have introduced an approach to cross-phase modulation based on a very-far-off-resonance single-photon detuning in a ladder system. By coupling the atomic ensemble, and the control and probe beams to an optical resonator the maximum cross-phase modulation is shown to be increased, compared to the optimal one-pass non-cavity case, by a factor proportional to the finesse of the resonator. A full classical description of the coherent fields inside the resonator is presented and no saturation of the atomic medium is required. This system is independent of the on-resonance optical depth of the atomic ensemble and it is self-consistent to an effective cooperativity of unity in the optimal regime of maximum XPM.  

Our protocol is expected to have a fast response to the nonlinear interaction due to the very-far-off-resonance condition, nevertheless the speed of the full operation is determined by the bandwidth of the resonator. Also, no self-modulation is induced onto the probe beam and no considerable group-velocity-mismatch between both beams is present. Even though a high finesse resonator would offer better XPM performance, the strong-cavity regime of cavity QED is not required in our protocol. We also note that the protocol shall not be confused with the dispersive regime of cavity QED.

The calculations are presented for Doppler-free expressions of the linear and non-linear susceptibilities, indicating that a direct experimental test of our results requires a laser-cooled atomic ensemble. Nevertheless due to the necessary high single-photon detuning this protocol could work similarly at room temperature if the single-photon detuning is larger than the Doppler broadening of the linear absorption profile, and the $F/4\pi$ enhancement is expected to hold. The Doppler-free condition for the nonlinear susceptibility at room temperature would be satisfied due to the resonator-coupling of both fields if $\lambda_c\approx\lambda_p$. Each component of the control beam inside the resonator would get either red-shifted or blue-shifted in the atom's frame of reference depending on the direction of the atom's velocity. Interaction only with counter-propagating components of the probe beam inside the resonator would induce a XP since the sum of the two frequencies would still satisfy the required two-photon detuning.

%A possible decrement in the XP for this case could be caused by the different spatial coupling strength due to the atoms flying through the intracavity standing waves. 

Linear absorption of the control beam is quite high ($\sim75\%$), nevertheless we have shown that transmission can be set close to 100$\%$ without paying much of a price in XPM. The process is also limited by two-photon absorption (TPA), which we have ignored in our analysis, but we note that for the optimal detuning parameters $\left|\mbox{Im}(\chi^{(3)})\right|\approx\left|\mbox{Re}(\chi^{(3)})\right|$. If strong transverse confinement (high intensity) of both beams is used, a larger two-photon detuning would be required to avoid TPA. 

We hope our results call for an experimental demonstration. Extending this protocol to the single-photon level will offer richer dynamics and will examine the ultimate advantages of the approach.

\section*{Acknowledgments}
J. M.-R. gratefully acknowledges helpful discussions with Curtis Broadbent, Chris Mullarkey, Gerardo Viza, Vivek Venkataraman, Alexander Gaeta, and Mark Kasevich; as well as the careful editing of Bethany Little. This work was supported by the DARPA-DSO grant number W31P4Q-12-1-0015.

%\bibliography{Paper-XPMinResonator2ArXiv-AfterReferees}

%

\end{document}